\documentclass{article}

\usepackage{microtype}
\usepackage{graphicx}
\usepackage{subcaption}
\usepackage{booktabs} 

\usepackage{hyperref}

\usepackage{tabularx}
\usepackage{booktabs}
\usepackage{array}

\usepackage[preprint]{icml2026}

\usepackage{amsmath}
\usepackage{amssymb}
\usepackage{mathtools}
\usepackage{amsthm}

\usepackage[capitalize,noabbrev]{cleveref}

\usepackage[textsize=tiny]{todonotes}

\newif\ifanonym
\anonymfalse  

\ifanonym
  \newcommand{\modelname}{Model X}
  \newcommand{\companyname}{Anonymous Institution}
  \newcommand{\delliba}{DEL A}
  \newcommand{\dellibb}{DEL B}
  \newcommand{\strelka}{DEL-CLS}
\else
  \newcommand{\modelname}{Hermes}
  \newcommand{\companyname}{Leash Biosciences, Inc.}
  \newcommand{\delliba}{Kin0}
  \newcommand{\dellibb}{AMA020}
  \newcommand{\strelka}{STRELKA}
\fi

\icmltitlerunning{\modelname{}: Large DEL Datasets Train Generalizable Protein-Ligand Binding Prediction Models}

\begin{document}

\twocolumn[
  \icmltitle{\modelname{}: Large DEL Datasets Train Generalizable Protein-Ligand Binding Prediction Models}



  \icmlsetsymbol{equal}{*}

  \begin{icmlauthorlist}
    \icmlauthor{Maxwell Kleinsasser}{lsh}
    \icmlauthor{Brayden J. Halverson}{lsh}
    \icmlauthor{Edward Kraft}{lsh}
    \icmlauthor{Sean Francis-Lyon}{lsh}
    \icmlauthor{Sarah E. Hugo}{lsh}
    \icmlauthor{Mackenzie R. Roman}{lsh}
    \icmlauthor{Ben Miller}{lsh}
    \icmlauthor{Andrew D. Blevins}{lsh}
    \icmlauthor{Ian K. Quigley}{lsh}
  \end{icmlauthorlist}
  
  \icmlaffiliation{lsh}{\companyname{}, Salt Lake City, Utah}

  \icmlcorrespondingauthor{Maxwell Kleinsasser}{max@leash.bio}
  \icmlcorrespondingauthor{Andrew D. Blevins}{andrew@leash.bio}
  \icmlcorrespondingauthor{Ian K. Quigley}{ian@leash.bio}

  \icmlkeywords{Machine Learning, ICML}

  \vskip 0.3in
]



\printAffiliationsAndNotice{}  

\begin{abstract}
The quality and consistency of training data remain critical bottlenecks for protein-ligand binding prediction. Public affinity datasets, aggregated from thousands of labs and assay formats, introduce biases that limit model generalization and complicate evaluation. DNA-encoded chemical libraries (DELs) offer a potential solution: unified experimental protocols generating massive binding datasets across diverse chemical and protein target space. We present \modelname{}, a lightweight transformer trained exclusively on DEL data from screens against hundreds of protein targets, representing one of the largest and most protein-diverse DEL training sets applied to protein-ligand interaction (PLI) modeling to date. Despite never seeing traditional affinity measurements during training, \modelname{} generalizes to held-out targets, novel chemical scaffolds, and external benchmarks derived from public binding data and high-throughput screens. Our results demonstrate that DEL data alone captures transferable protein-ligand interaction representations, while \modelname{}'s minimal architecture enables inference speeds suitable for large-scale virtual screening.
\end{abstract}

\section{Introduction}

Accurately modeling PLI is a foundational challenge in drug discovery. Recently, substantial progress has been made in biological complex structure prediction, most notably through AlphaFold3 and its open-source adaptations \citep{abramson2024alphafold3, wohlwend2024boltz1, chaidiscovery2024chai1}. However, it is important to distinguish structure prediction---a generative modeling task---from PLI binding prediction: the task of determining whether and how strongly a given protein-ligand pair will interact. Binding prediction is typically framed as either regression over experimental affinity values (e.g., IC$_{50}$, K$_d$), often called protein-ligand scoring, or binary classification over bind/no-bind labels.

In contrast to crystallography and sequence-based data in structural studies, data quality remains a pervasive challenge for PLI modeling. Large repositories such as BindingDB \citep{liu2007bindingdb} and ChEMBL \citep{gaulton2012chembl} aggregate millions of affinity measurements curated from published articles and patents. However, these measurements originate from thousands of different labs, assays, and experimental protocols, resulting in data that is notoriously difficult to standardize and riddled with biases \citep{kramer2012bioactivity, harren2023latentbias, volkov2022frustration, blevins2025cleverhans}. Datasets that do employ consistent experimental protocols are generally too limited in their coverage of protein or chemical space to train generalizable models \citep{davis2011kinase, metz2011kinome}.

Despite these training data challenges, effective computational models for PLI binding prediction do exist. Classical physics-based methods have long provided the foundation for affinity estimation, from rigorous free energy perturbation (FEP) calculations to faster endpoint approximations like MM-PBSA and MM-GBSA \citep{wang2015fep, genheden2015mmpbsa}. However, these approaches remain too computationally demanding for large-scale virtual screening, motivating the development of machine learning alternatives. ML approaches have proliferated over the past two decades, spanning architectures from random forests \citep{ballester2010mlsf} to deep learning models including sequence-based methods \citep{ozturk2018deepdta}, graph neural networks \citep{nguyen2021graphdta}, and structure-based approaches such as Boltz-2, which trains binding affinity and binary classification heads on top of a pretrained AlphaFold3-like architecture \citep{passaro2025boltz2}.

Evaluating these approaches is equally challenging. Few high-quality benchmark datasets exist because they inherit the same problems as training data. Data leakage is particularly problematic: it is pervasive, difficult to detect, and lacks standardized mitigation strategies \citep{graber2025cleansplit}. Curated benchmarks with stronger leakage protections (temporal, assay, or protein-family splits) tend to be too small for comprehensive evaluation \citep{gilson2025casp16}. DELs \citep{brenner1992encoded, gironda2021delreview} offer a potential solution by unifying experimental protocols across protein targets while exploring vast regions of chemical space in single experiments.

DELs are libraries of small molecules where each compound is covalently attached to a unique DNA barcode, enabling the synthesis and screening of up to billions of compounds in a single experiment. In a DEL screen, the entire library can be incubated with an immobilized protein target, non-binders are washed away, and the remaining compounds are identified by sequencing their DNA tags; enrichment over background indicates binding. This massively parallel approach generates binding data at a scale orders of magnitude larger than traditional high-throughput screens, though the resulting enrichment scores are noisy proxies for true binding affinity.

The massive scale of DEL data makes it well-suited for training machine learning models. Several DEL datasets have been publicly released \citep{quigley2024belka, lim2024kindel, iqbal2025delevaluation}, and multiple groups have reported success training ML models on this data \citep{mccloskey2020del, iqbal2025delevaluation, lim2024kindel}. However, publicly available DEL datasets are severely limited in protein diversity, restricting prior work to single-protein models. While these models have demonstrated experimental validation of virtual screening hits, such validation has been confined to small numbers of compounds against the same protein target used for training. What remains to be tested is whether PLI representations learned from DEL data can transfer more broadly: across held-out protein targets, unseen chemical scaffolds, and binding measurements from entirely different experimental systems.

In this work, we present \modelname{}: a fast, sequence-only PLI binding prediction model trained exclusively on binarized DEL screening data across hundreds of unique protein targets. We find that DEL-only training produces PLI representations that generalize to held-out protein targets, novel chemical scaffolds, and external benchmarks derived from non-DEL affinity measurements. \modelname{}'s minimal architecture enables inference speeds multiple orders of magnitude faster than state-of-the-art structure-based PLI models, though with a corresponding trade-off in predictive performance.

\section{Methods}

\begin{figure}[t]
  \centering
  \includegraphics[width=\columnwidth]{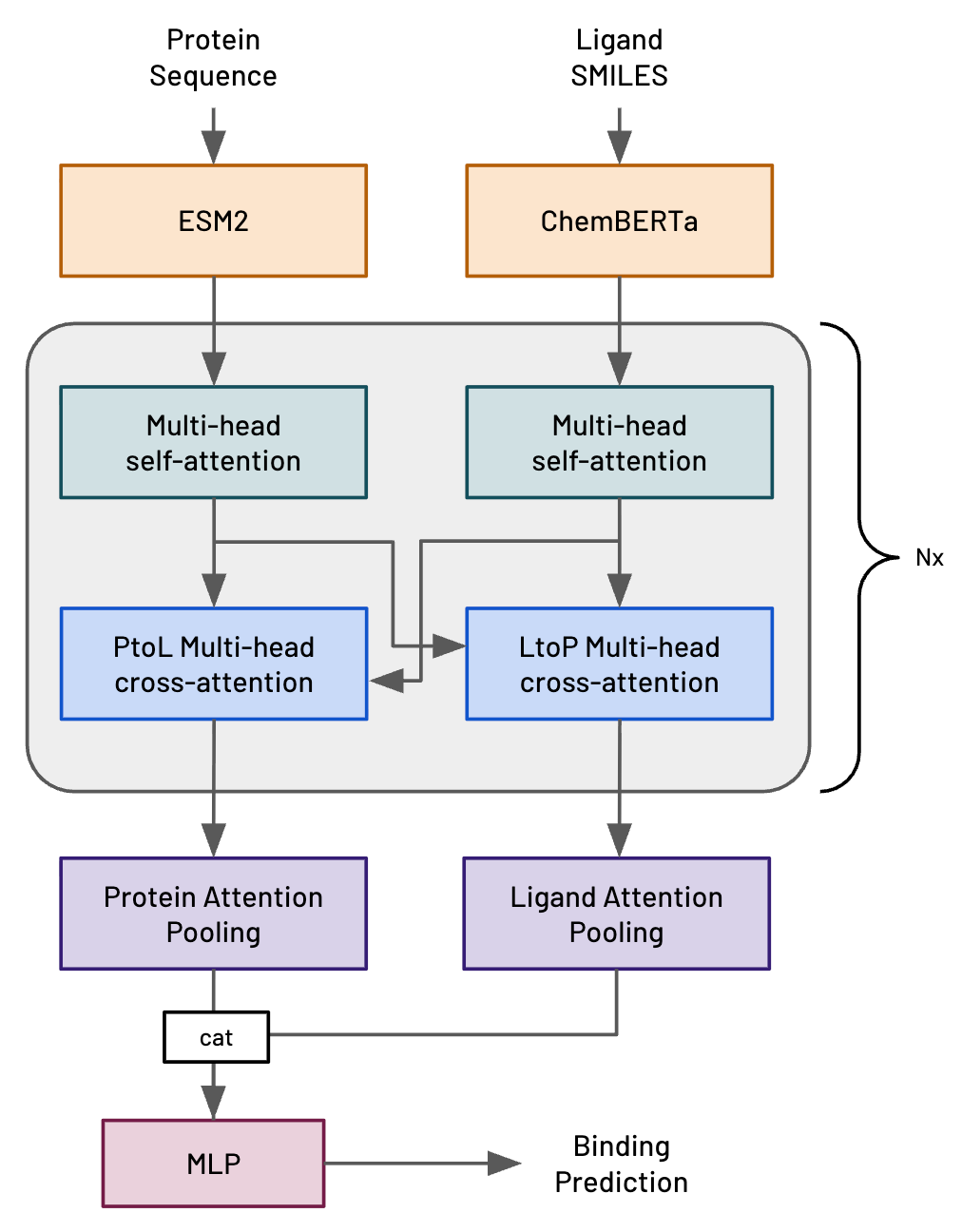}
  \caption{\textbf{\modelname{} architecture diagram.}}
  \label{fig:hermes-architecture}
\end{figure}

\begin{figure*}[t]
  \centering
  \includegraphics[width=\textwidth]{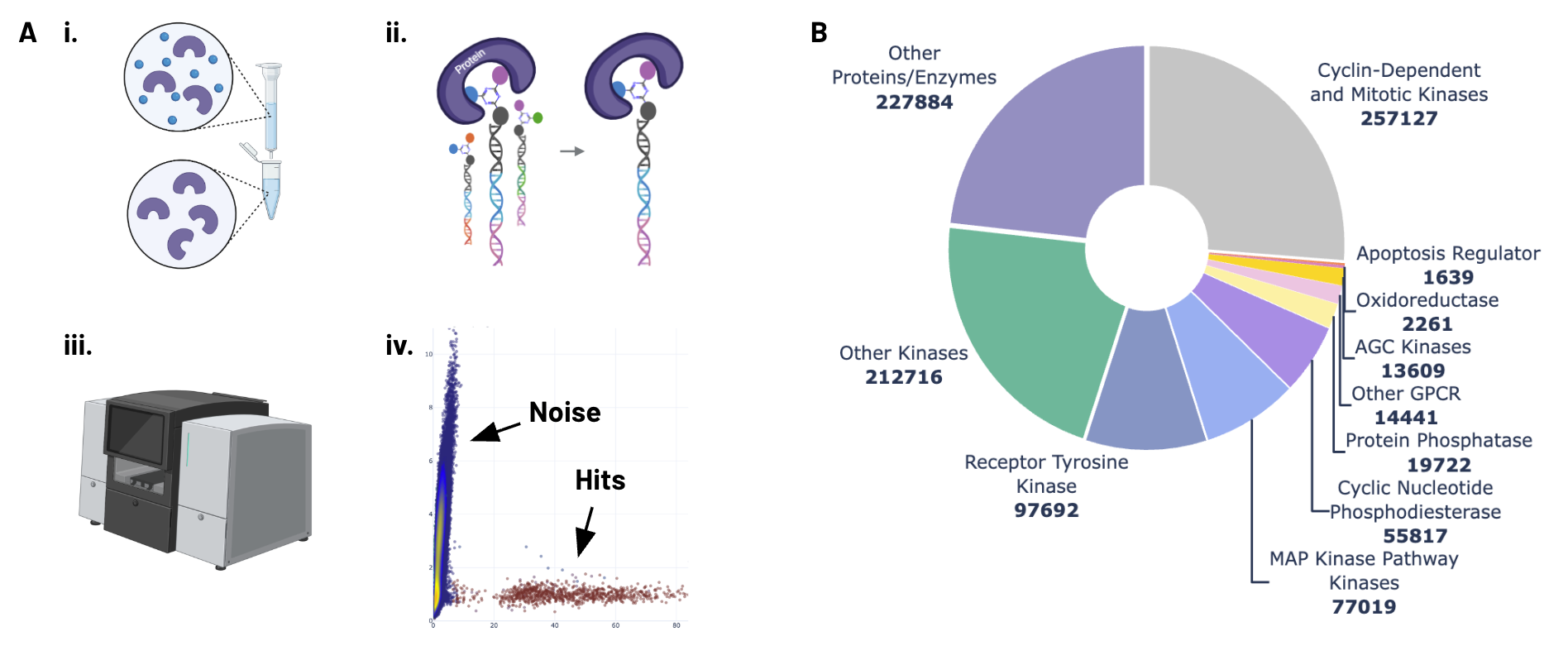}
  \caption{\textbf{DEL screening workflow and training data composition.} (A) Schematic of the DEL screening assay: (i) protein expression, purification, and immobilization; (ii) library incubation with iterative washing to remove non-binders; (iii) PCR amplification and sequencing of retained compounds; (iv) quality control, enrichment quantification, and hit classification. (B) Distribution of training samples across protein targets, colored by protein family.}
  \label{fig:training-data}
\end{figure*}

\subsection{\modelname{}}

\modelname{} employs a lightweight transformer-based architecture for binary protein-ligand binding prediction. The model leverages pre-trained sequence embedding models, ESM2 \citep{lin2023esm2} for protein sequences and ChemBERTa \citep{chithrananda2020chemberta} for ligand SMILES strings, followed by a joint cross-attention module, attention pooling layers, and an MLP that outputs a scalar binding probability.

The joint cross-attention module \citep{vaswani2017attention} interleaves standard self-attention blocks, which operate on protein and ligand sequence representations independently, with cross-attention blocks that facilitate information flow between protein sequence tokens and ligand tokens. Two attention pooling layers aggregate the protein and ligand sequence representations into fixed-length vectors. Each attention pooling layer learns a scoring function that maps token embeddings to scalar attention weights; the softmax of these scores determines each token's contribution to the pooled sequence-level representation. The pooled protein and ligand representations are concatenated and passed through an MLP to produce a final predicted binding probability (\cref{fig:hermes-architecture}).

We observe significantly improved performance when aggregating predictions from multiple \modelname{} checkpoints trained with varying training set sampling strategies and hyperparameter configurations. Evaluation metrics presented in this work are computed using the mean predictions of 9 \modelname{} checkpoints (\cref{sec:checkpoint-details}). Checkpoints were selected based on performance on internal building block splits of the training set as well as a curated validation set of publicly available binding data (\cref{sec:evaluation}).

\subsection{Training Data}
\label{sec:training-data}

\modelname{} is trained exclusively on protein-ligand interaction data derived from DEL screens against a 6.5 million member chemical library referred to as ``\delliba{}.'' \delliba{} was constructed as a 3-cycle library with 38 cores connected to 384 and 446 building blocks. Building block selection was performed using a Thompson sampling scheme to optimize drug-like properties and predicted likelihood of binding several kinase targets \citep{klarich2024thompson}.

Individual models were trained on data from up to 239 unique proteins, approximately two-thirds of which were kinases.

DEL data is binarized using a hit-calling procedure that detects significant enrichment over control screens. Control screens include: (1) DEL-only screens, which measure the baseline relative proportions of library members in the unscreened DEL; (2) bead-only or no-target controls, which measure relative proportions of library members processed through the DEL assay in the absence of a protein target; and (3) additional proprietary controls. All protein target screens undergo manual review against control screens for sufficient screening depth and visual presence of enriched compounds.

To be included in the training data pool as a positive sample, sequencing counts for individual molecules must significantly exceed those expected from unenriched compounds relative to the DEL-only control. Conversely, compounds with counts confidently below those expected under a low enrichment factor are included as negative samples. Compounds not confidently classified as enriched or unenriched relative to DEL-only controls are excluded from the training pool, as are compounds enriched relative to matched control screens.

Because the vast majority of compounds do not bind any given protein target, the resulting training pool from hit-calling is highly imbalanced such that training on the full dataset is unfeasible and undesirable. To address this imbalance, we sample and stratify the training pool. The number of positive samples contributed from a single protein is capped, retaining hits with the highest sequencing counts in the corresponding screen, which prevents individual protein targets from having outsized representation in the final training set. We then sample a fixed number of negatives per positive per protein. Per-protein negatives are drawn from pools of random negatives as well as pools of hard negatives designed to prevent memorization during training. This sampling process defines a set of parameters that strongly influences training outcomes and is varied across training runs to produce diverse model checkpoints.

\subsection{Evaluation}
\label{sec:evaluation}

\begin{table*}[t]
\centering
\caption{\textbf{Training and evaluation dataset statistics.}}
\label{tab:dataset_details}
\begin{tabularx}{\textwidth}{l r r r >{\raggedright\arraybackslash}X}
\toprule
Dataset & \#Targets & \#Pos & \#Neg & Description \\
\midrule
DEL Training Set & 239 & 146{,}854$^{*}$ & 0.83M$^{*}$ & DEL training data, ``\delliba{}'' molecule library, enriched molecules as positives, unenriched as negatives \\
DEL Protein Split & 164 & 28{,}710 & 0.17M & DEL test data, ``\delliba{}'' molecule library, proteins not in training set \\
DEL Chemical Library Split & 59 & 27{,}585 & 3.02M & DEL test data, unseen library ``\dellibb{}'', proteins in training set \\
Public Binders/Decoys & 403 & 30{,}372 & 2.82M & Papyrus++ binders as positives, property-matched synthetic decoys as negatives \\
MF-PCBA & 26 & 13{,}480 & 7.68M & PubChem BioAssay confirmed HTS binders as positives, HTS inactives as negatives \\
\bottomrule
\end{tabularx}
\footnotesize
$^{*}$Training set statistics represent the most common configuration among the 9 model checkpoints; the negative-to-positive sampling ratio is varied across training runs.
\end{table*}

We evaluate \modelname{} on a binary binding classification task across a diverse set of benchmarks designed to test generalization at multiple levels, including unseen proteins, assays, and chemical space. We report per-protein AUROC and average precision (AP) scores. Reporting per-protein rather than aggregate rank-based metrics is common practice, as mixing predictions across protein targets unnecessarily dilutes performance across all PLI binding prediction models. We further stratify results by protein family (kinase vs. non-kinase) to assess whether the kinase-enriched training set composition translates to differential generalization performance. We initially present \modelname{} performance in isolation for several reasons. Primarily, we seek to emphasize generalization from DEL-only training data as the central contribution of this work. Furthermore, a substantial portion of our evaluation datasets are contained in the training sets of virtually all models we might benchmark against, making rigorous and fair comparisons difficult and potentially misleading. Moreover, we see wide variation in model performance depending on which benchmark sets we use. The internal benchmarks in this work were selected before any inference runs to maintain rigor. We provide performance comparisons with the state-of-the-art PLI model Boltz-2 and an XGBoost baseline in \cref{sec:benchmark-results}.

The \textbf{DEL Protein Split} benchmark comprises DEL data screened with the same \delliba{} molecule library as the \modelname{} training set, but against protein targets from DEL screens not seen during training. The protein family distribution reflects that of the training set, including many similar targets: just over half of the protein targets are kinases, with the remainder representing a diverse set of human proteins selected for drug discovery relevance. This dataset evaluates ``cold target'' generalization within the same assay and chemical space, as well as the broad reliability of DEL screening data.

The \textbf{DEL Chemical Library Split} benchmark (\strelka{}) comprises DEL data from a 1 million member chemical library (``\dellibb{}'') screened against protein targets predominantly contained in the \modelname{} training set. This dataset evaluates ``cold ligand'' generalization within the same assay and protein target space.

The \textbf{Public Binders/Decoys} dataset is derived from publicly available experimental binding affinity data and carefully curated synthetic decoys. Positive samples are drawn from Papyrus++ \citep{bequilleux2022papyrus}, a curated dataset of experimental binding affinity measurements from ChEMBL and other sources. We define ``active'' compounds as Papyrus++ molecules within a specified range of molecular properties and a pChEMBL value greater than 7 (corresponding to sub-100 nM potency). For negative samples, we use molecular-property-matched decoys from GuacaMol \citep{brown2019guacamol}, a collection of drug-like ChEMBL molecules designed for de novo molecular generation benchmarking.

Selecting decoys from GuacaMol rather than known binders of other protein targets is intended to mitigate false negatives arising from unevaluated compound promiscuity \citep{vogel2011dekois}. We acknowledge the limitations and implicit biases inherent in synthetic decoy curation \citep{bauer2013dekois2}. However, we find that selecting negative samples from known low-affinity compounds evaluated against the protein target of interest yields benchmarks that more strongly measure a model's adherence to the implicit biases of public binding datasets \citep{harren2023latentbias} rather than true PLI signal. To further reduce false negatives among decoys, we enforce that no decoy has a Tanimoto similarity greater than 0.3 to any binder for its protein. This dataset serves as the external validation source for model checkpoint selection. To mitigate model-selection overfitting, we split this data into validation and test sets by protein, using the validation set for checkpoint selection and reporting only test set performance here.

The \textbf{MF-PCBA} dataset is derived from MF-PCBA (Multifidelity PubChem BioAssay) \citep{tran2023mfpcba}, a curated dataset of high-throughput screening assays requiring confirmatory dose-response data for binders. This dataset avoids some biases of large public binding affinity datasets and synthetic decoy-based benchmarks because negatives are drawn from primary HTS screens. However, it is constrained in protein target diversity and includes proteins outside typical human drug discovery efforts. Furthermore, its composite assay data is highly heterogeneous with respect to methods used for qualifying ``active'' compounds. The reliance on a variety of indirect and semi-direct biochemical assays is subject to high error rates \citep{tiikkainen2013errorrates}. We preprocess this dataset by manually excluding assays with phenotypic readouts and removing molecules flagged as pan-assay interference compounds (PAINS) \citep{baell2010pains} via RDKit.

\begin{figure*}[!t]
  \centering
  \includegraphics[width=\textwidth]{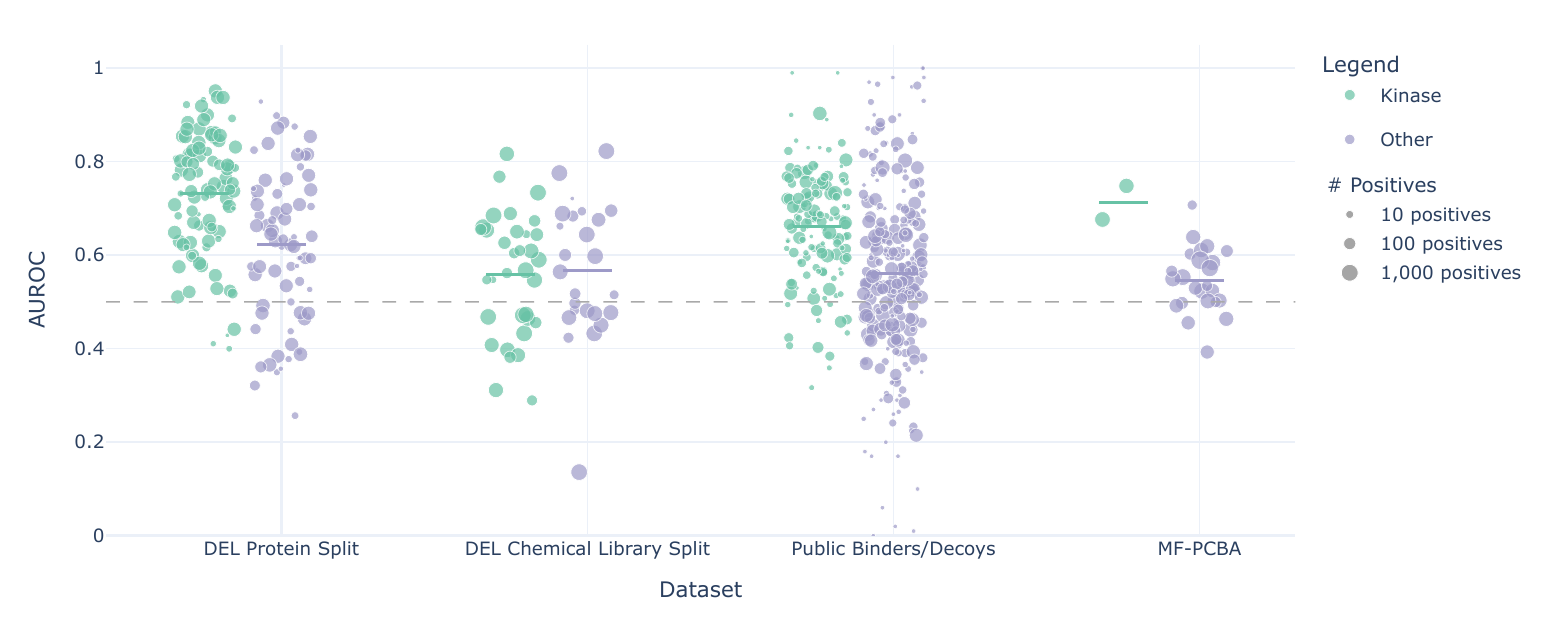}
  \caption{\textbf{\modelname{} per-protein AUROC scores by evaluation dataset.} Point size indicates the number of positive samples for each protein target. Results are stratified by protein family (kinase vs. non-kinase) to assess whether the kinase-enriched training set composition translates to differential evaluation performance. The dashed line indicates random classifier performance (AUROC = 0.5). Solid horizontal lines indicate mean AUROC within each dataset and protein family group.}
  \label{fig:results_scatter}
\end{figure*}

\begin{table*}[!t]
\centering
\caption{\textbf{\modelname{} per-protein performance summary.} Scores reported as mean and standard deviation per-protein AUROC and Average Precision scores.}
\label{tab:hermes_performance}
\begin{tabular}{llrccc}
\toprule
Dataset & Family & N & Mean AUROC & Mean AP & Pos Rate \\
\midrule
DEL Protein Split & Kinase & 91 & $0.7325 \pm 0.1284$ & $0.3225 \pm 0.1302$ & $0.1372 \pm 0.0100$ \\
 & Other & 73 & $0.6227 \pm 0.1603$ & $0.2350 \pm 0.1102$ & $0.1362 \pm 0.0116$ \\
 & All & 164 & $0.6837 \pm 0.1535$ & $0.2835 \pm 0.1292$ & $0.1368 \pm 0.0108$ \\
\midrule
DEL Chemical Library Split & Kinase & 34 & $0.5596 \pm 0.1276$ & $0.0199 \pm 0.0206$ & $0.0096 \pm 0.0004$ \\
 & Other & 25 & $0.5672 \pm 0.1430$ & $0.0201 \pm 0.0222$ & $0.0087 \pm 0.0013$ \\
 & All & 59 & $0.5628 \pm 0.1344$ & $0.0200 \pm 0.0213$ & $0.0092 \pm 0.0010$ \\
\midrule
Public Binders/Decoys & Kinase & 139 & $0.6620 \pm 0.1200$ & $0.0366 \pm 0.0731$ & $0.0100 \pm 0.0009$ \\
 & Other & 264 & $0.5608 \pm 0.1866$ & $0.0331 \pm 0.0930$ & $0.0105 \pm 0.0018$ \\
 & All & 403 & $0.5957 \pm 0.1735$ & $0.0343 \pm 0.0867$ & $0.0103 \pm 0.0015$ \\
\midrule
MF-PCBA & Kinase & 2 & $0.7123 \pm 0.0359$ & $0.0075 \pm 0.0014$ & $0.0020 \pm 0.0001$ \\
 & Other & 24 & $0.5467 \pm 0.0667$ & $0.0023 \pm 0.0018$ & $0.0018 \pm 0.0013$ \\
 & All & 26 & $0.5594 \pm 0.0784$ & $0.0027 \pm 0.0022$ & $0.0018 \pm 0.0013$ \\
\bottomrule
\end{tabular}
\end{table*}

\subsection{Benchmark Models}
\label{sec:benchmark-models}
We contextualize our results by comparing \modelname{} performance against two baseline models: a state-of-the-art structure-based deep learning model, Boltz-2, and a memorization-based baseline. Importantly, we include benchmark model performance as context revealing the nuances of evaluating PLI models, we do not claim state-of-the-art task performance in this work.

\paragraph{Boltz-2.}
Boltz-2 is the current state-of-the-art model for protein-ligand binding prediction \citep{passaro2025boltz2}. Boltz-2 trains both binding affinity regression and binary binding classification heads on top of a pre-trained AlphaFold3-like structure prediction architecture, representing a substantially more expressive and compute-intensive model class than \modelname{}. We used Boltz-2 version 2.1.1 with 5 recycling steps and otherwise default inference parameters. Inference was performed on clusters of 8$\times$ NVIDIA H200 GPUs. Notably, this evaluation likely represents one of the largest and most leak-proof assessments of Boltz-2 performed to date.

\paragraph{XGBoost Baseline.}
We train an XGBoost classifier as a baseline designed to evaluate performance achievable primarily through DEL training set memorization. We train this model using the same DEL training data as the \modelname{} ensemble. Protein and ligand inputs are represented by concatenated EMS2-650M embeddings and ECFP4 fingerprints. Results reported here represent the best model of a hyperparameter search based on the mean evaluation set scores.

\paragraph{Benchmark Evaluation Data.} We evaluate on subsets of the DEL Protein Split, DEL Chemical Library Split, and MF-PCBA datasets, each subsampled to approximately 50,000 total samples to make Boltz-2 inference computationally feasible. We additionally exclude proteins with sequence lengths greater than 1024. We do not include Boltz-2 evaluations on the Public Binders/Decoys dataset, as these data represent a substantial portion of Boltz-2's training set. Furthermore, some portion of the MF-PCBA benchmark is included in Boltz-2's training set \citep{passaro2025boltz2}. The vast majority of protein targets in the DEL benchmarks will have appeared in the Boltz-2 training set as well. We also report XGBoost Baseline performance on the full Public Binders/Decoys dataset.

\section{Results}

Overall, performance varies substantially across protein targets (\cref{fig:results_scatter}), but is strongest on the DEL Protein Split benchmark, which is likely primarily driven by similarity to the training set in both assay format and chemical space (see \cref{sec:tanimoto-similarity}). Performance on the Public Binders/Decoys dataset, the most comprehensive orthogonal evaluation set, demonstrates that PLI representations learned from DEL data transfer strongly within kinases, and modestly otherwise. Improved performance among kinases is  apparent in three of four evaluation sets and consistent with the kinase-enriched composition of the training data. The effect is especially pronounced on MF-PCBA, where the only two kinases (GSK3A, GSK3B) substantially outperform the otherwise non-standard protein targets in the dataset. This result suggests that targeted training data generation improves generalization performance within protein classes.

Per-protein performance varies substantially, particularly among protein targets with few positive samples and non-kinases. We attribute this variance in part to the inherent sparsity of PLI datasets: positive samples for any given target tend to over-represent particular chemical scaffolds, limiting the diversity of the positive class even for well-sampled targets.

Performance on DEL Chemical Library Split is low for all models. Average \modelname{} AUROC scores are below those on Public Binders/Decoys despite the shared DEL assay format and slightly greater similarity to the training set (see \cref{sec:tanimoto-similarity}). Whether this reflects model limitations or characteristics of the data is unclear. Notably, Boltz-2 performance achieves near-perfect ranking on this evaluation set for one protein, BRD4, known binders of which are both publicly available and were used to construct the library used for the DEL Chemical Library Split (see \cref{sec:strelka}).

\subsection{Benchmark Comparisons}
\label{sec:benchmark-results}

\begin{figure*}[!htbp]
  \centering
  \ifanonym
    \includegraphics[width=\textwidth]{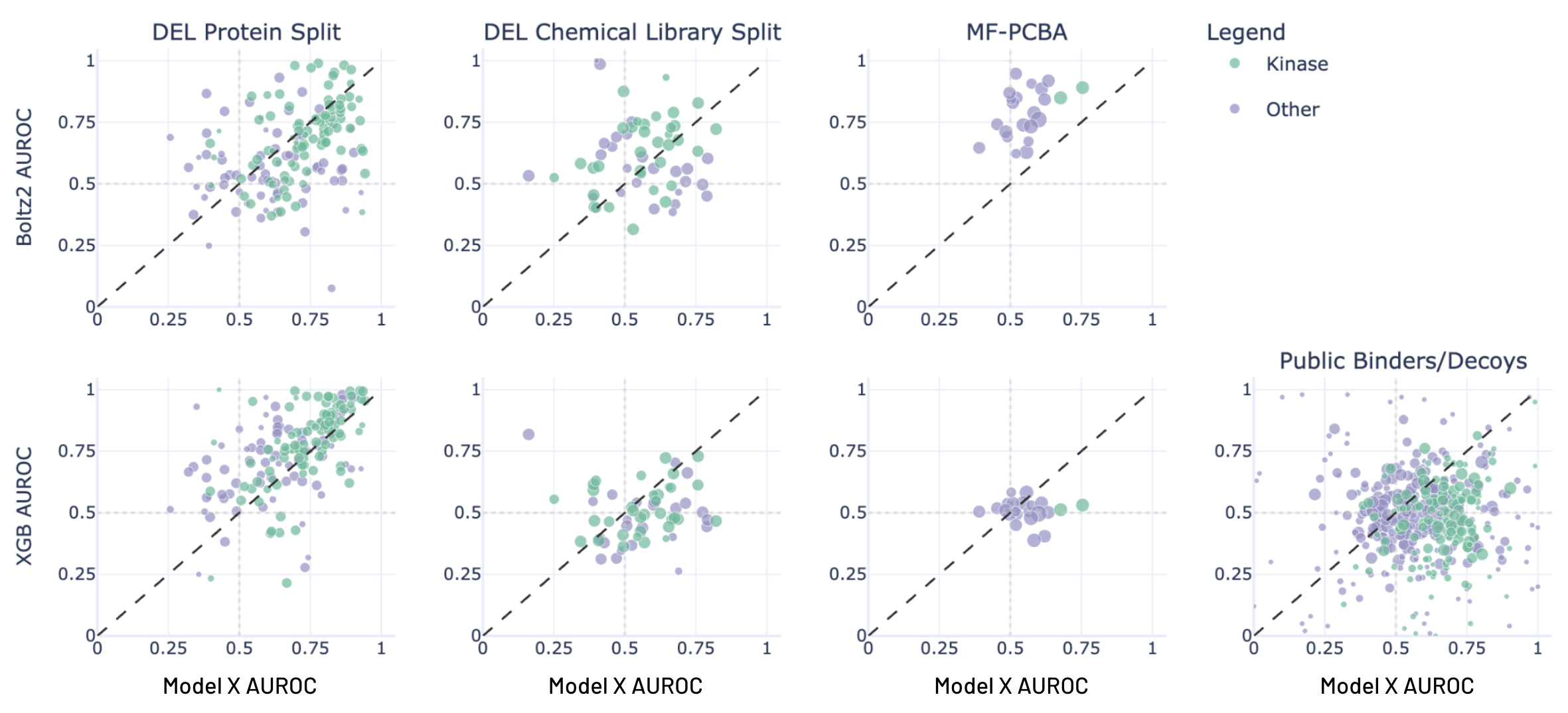}
  \else
    \includegraphics[width=\textwidth]{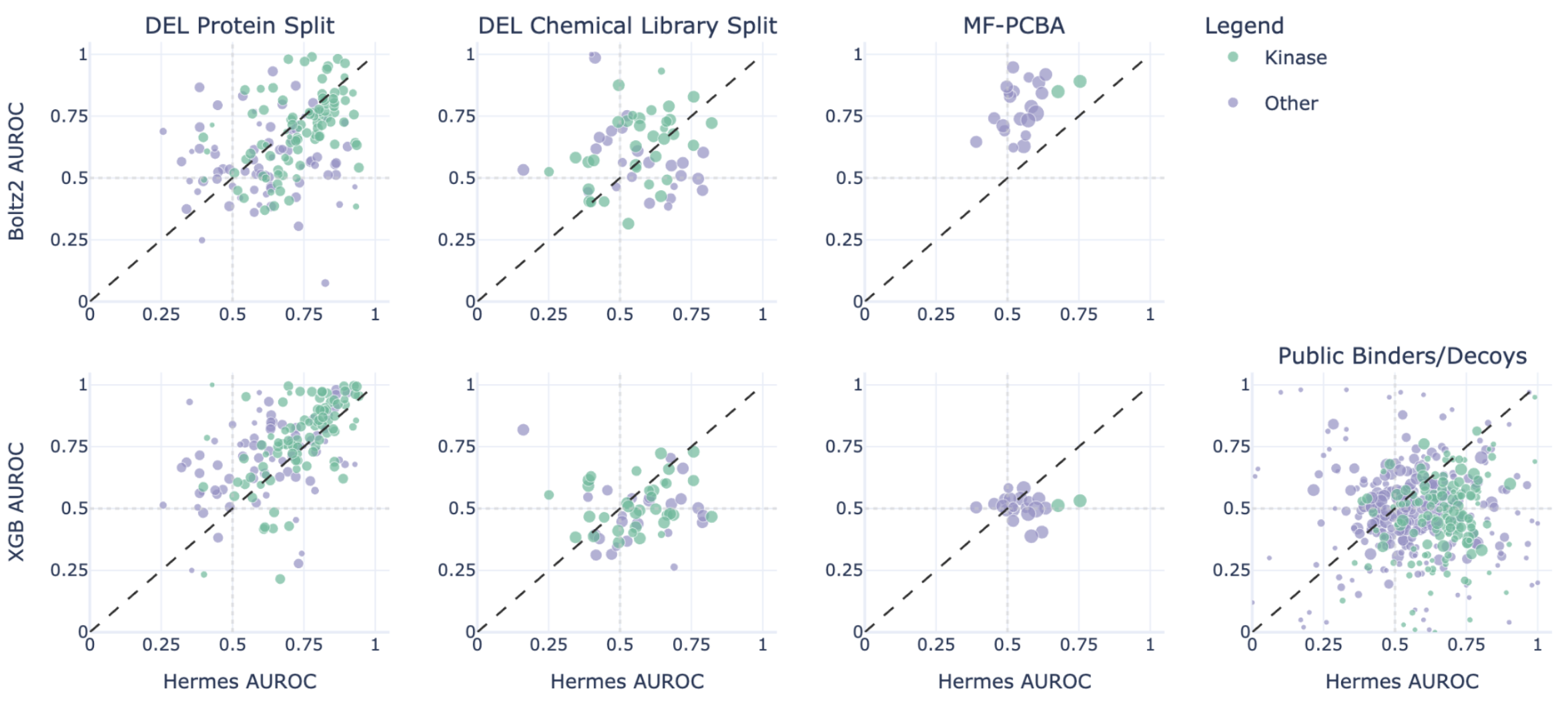}
  \fi
  \caption{\textbf{\modelname{} vs. benchmarks per-protein AUROC comparison across evaluation datasets.} Point size indicates number of binders within the protein target. All but the Public Binders/Decoys dataset are subsampled to ~50k samples for Boltz-2 inference time/cost feasibility.}
  \label{fig:Boltz-2-comparison}

  \vspace{4em}

  \centering
  \captionof{table}{\textbf{\modelname{} vs benchmarks per-protein AUROC comparison.} AUROC columns indicate mean + standard deviation AUROC scores across all targets in the evaluation set. }
  \label{tab:hermes_vs_benchmarks}
  \begin{tabular*}{\textwidth}{@{\extracolsep{\fill}}llrccc@{}}
  \toprule
  Dataset & Family & N & XGB Baseline AUROC & \modelname{} AUROC & Boltz-2 AUROC \\
  \midrule
  DEL Protein Split & Kinase & 91 & $0.779 \pm 0.168$ & $0.734 \pm 0.128$ & $0.693 \pm 0.151$ \\
   & Other & 73 & $0.703 \pm 0.159$ & $0.619 \pm 0.163$ & $0.571 \pm 0.152$ \\
   & All & 164 & $0.745 \pm 0.168$ & $0.683 \pm 0.155$ & $0.639 \pm 0.163$ \\
  \midrule
  DEL Chemical Library Split & Kinase & 34 & $0.517 \pm 0.099$ & $0.564 \pm 0.128$ & $0.627 \pm 0.147$ \\
   & Other & 25 & $0.485 \pm 0.127$ & $0.559 \pm 0.151$ & $0.583 \pm 0.154$ \\
   & All & 59 & $0.504 \pm 0.113$ & $0.562 \pm 0.138$ & $0.608 \pm 0.152$ \\
  \midrule
  MF-PCBA & Kinase & 2 & $0.522 \pm 0.009$ & $0.716 \pm 0.038$ & $0.870 \pm 0.021$ \\
   & Other & 20 & $0.505 \pm 0.048$ & $0.538 \pm 0.059$ & $0.781 \pm 0.098$ \\
   & All & 22 & $0.507 \pm 0.046$ & $0.554 \pm 0.077$ & $0.789 \pm 0.097$ \\
  \midrule
  Public Binders/Decoys & Kinase & 139 & $0.461 \pm 0.168$ & $0.662 \pm 0.120$ & — \\
   & Other & 264 & $0.493 \pm 0.186$ & $0.561 \pm 0.187$ & — \\
   & All & 403 & $0.482 \pm 0.180$ & $0.596 \pm 0.173$ & — \\
  \bottomrule
  \end{tabular*}
\end{figure*}

\cref{fig:Boltz-2-comparison} and \cref{tab:hermes_vs_benchmarks} compare \modelname{} classification performance with Boltz-2 and the XGBoost Benchmark model.

\paragraph{Boltz-2.} On the DEL Protein Split, both \modelname{} and the XGBoost Baseline outperform Boltz-2 across both kinase and non-kinase targets. Given strong performance from the XGBoost benchmark model, we attribute this result primarily to the DEL training data from the same assay system and chemical library, screened against proteins predominantly similar to those in the training set. However, that Boltz-2, trained on public data alone, achieves significant performance on this dataset is strong evidence of the quality of both the model and the dataset in capturing meaningful PLI signal.

On the DEL Chemical Library Split, Boltz-2 performs better but similarly poorly compared to \modelname{}. This suggests either that there is a data quality problem with the evaluation set or that the evaluation set is particularly challenging. We interpret Boltz-2's near-perfect performance on the BRD4 protein target as evidence for the latter; see \cref{sec:strelka} for more details. 

On the MF-PCBA dataset, Boltz-2 outperforms \modelname{} substantially, though with an important caveat that an unknown (at the time of writing) portion of this dataset is contained in the Boltz-2 training set.

\begin{table*}[h]
\centering
\caption{Inference speed comparison.}
\label{tab:inference-speed-main}
\begin{tabular}{lccc}
\toprule
Model & Hardware & Samples/sec/GPU & Relative Speedup \\
\midrule
\modelname{} Ensemble & H200 & 28.2 & $\sim$500--700$\times$ \\
Boltz-2 & H100 & $\sim$0.04 & 1$\times$ \\
\bottomrule
\end{tabular}
\end{table*}

\paragraph{XGBoost Baseline.} The XGBoost Baseline model performances are centered around random AUROC scores for all but the DEL Protein Split, on which it achieves the top score of 0.745 average AUROC. The efficacy of memorized protein-ligand features on this evaluation set diminishes its value as a benchmark dataset for \modelname{} against other affinity models with different training sets. This performance is driven largely by overlapping hit profiles of targets similar to training set proteins, see \cref{sec:benchmark-details}. This result highlights the unique challenges of PLI model evaluation on large-scale DEL data, but is also demonstrative of the density and reliability of enrichment in DEL screens among similar targets.

\paragraph{Inference Speed.}
Virtual screening applications require evaluating millions to billions of protein-ligand pairs, making inference speed a practical constraint on model selection. \modelname{}'s sequence-only architecture avoids the computational overhead of structure prediction and enables protein-embedding caching. We benchmark \modelname{} inference at 28.2 samples/second/GPU on H200, compared to Boltz-2's reported $\sim$0.04 samples/second/GPU on H100 \citep{passaro2025boltz2}. Accounting for hardware differences ($\sim$1.2--1.5$\times$ H200 vs. H100), this represents a $\sim$500--700$\times$ speedup. For large-scale virtual screening campaigns, this speedup may justify the performance gap on certain benchmarks, particularly for protein targets well-characterized in \modelname{}'s training data.

\section*{Discussion}

These results demonstrate the value of large DEL datasets for training generalizable PLI binding prediction models. The lightweight \modelname{} architecture is well-suited for virtual screening applications, though experimental validation of prospective predictions remains future work. Our results further suggest that targeted training data generation improves generalization within protein target classes, which could be especially valuable for protein families underrepresented in public binding affinity datasets \citep{oprea2018druggablegenome}.

We suspect \modelname{}'s predictive performance is limited by its architecture. Incorporating domain-specific inductive biases into architectures for biological complex modeling has proven highly effective in structure prediction \citep{jumper2021alphafold2}, and similar approaches are likely to benefit PLI binding prediction models, though at the cost of increased training and inference time. We anticipate meaningful gains through more sophisticated protein-ligand input representations, including structural augmentation of training data using co-folding models such as Boltz-2.

Training and evaluation dataset composition and pre-processing are also likely limitations of \modelname{} performance.  Cross-assay generalization is likely capped by label noise inherent in the DEL training data compounded with label noise in evaluation datasets derived from different experimental systems. Furthermore, the binarization of training data likely constrains model expressivity. Incorporating normalized sequencing counts from DEL screens to model relative binding strengths, rather than binary labels, remains a compelling direction for future work. Furthermore, despite significant generalization performance, we see that \modelname{} performance is strongly correlated with the memorization benchmark model on data similar to the training set, suggesting \modelname{} learning is still influenced by training data memorization, motivating future effort on training data sampling and filtering to combat the efficacy of memorization.

\section*{Conclusion}

We presented \modelname{}, a lightweight transformer for protein-ligand binding prediction trained exclusively on DEL screening data across hundreds of protein targets. Despite never training on traditional affinity measurements, \modelname{} generalizes to held-out protein targets, novel chemical scaffolds, and external benchmarks derived from public binding data and high-throughput screens. These results demonstrate that DEL data alone enables the learning of transferable protein-ligand interaction representations. As methods continue to improve and DEL datasets continue to substantially outpace public affinity data in growth, DEL-trained models are poised to drive the next generation of PLI prediction.

\bibliography{references}
\bibliographystyle{icml2026}

\newpage
\appendix
\onecolumn

\section{Checkpoint and Training Details}
\label{sec:checkpoint-details}

All checkpoints share the following training configuration. We use ChemBERTa-77M-MTR for ligand embeddings with all parameters trainable, and ESM2-150M for protein embeddings with the final 4 layers trainable. Dropout is set to 0.1 throughout. Models are trained with a batch size of 256 using the AdamW optimizer with weight decay 0.005. The learning rate warms up to $1 \times 10^{-5}$ over 500 steps, then decays to $1 \times 10^{-6}$ over 30K steps. We use focal loss with $\alpha = 0.75$ and $\gamma = 1.0$. \Cref{tab:checkpoint-details} summarizes the architectural and dataset variations across checkpoints.

\paragraph{Ablations.} Architectural ablations are not included in this work. We do not claim that our primary result is driven by a novel architecture. Rather, \modelname{} was designed to be a minimal deep-learning architecture potentially capable of nontrivial transfer, in order to facilitate fast and cost-effective experimentation on novel training data.

\begin{table}[H]
\centering
\caption{\textbf{Model checkpoint configurations and training details.} Hidden dim refers to the dimension of the token representations in the Joint Cross-Attention (Binding) Module. Binding module blocks consist of protein and ligand self-attention layers, a protein-to-ligand cross-attention layer, and a ligand-to-protein cross-attention layer (See \cref{fig:hermes-architecture}). }
\label{tab:checkpoint-details}
\small
\begin{tabular}{lrrrrrrrrr}
\toprule
 & \multicolumn{9}{c}{Checkpoint ID} \\
\cmidrule(lr){2-10}
 & 0 & 1 & 2 & 3 & 4 & 5 & 6 & 7 & 8 \\
\midrule
Hidden dim & 128 & 128 & 128 & 128 & 128 & 128 & 512 & 512 & 512 \\
Attention heads & 4 & 4 & 4 & 4 & 4 & 4 & 8 & 8 & 8 \\
Binding module blocks & 16 & 16 & 16 & 16 & 16 & 16 & 8 & 8 & 8 \\
Trainable parameters & 35M & 35M & 35M & 35M & 35M & 35M & 124M & 124M & 124M \\
Training samples & 980K & 818K & 583K & 725K & 583K & 622K & 622K & 1.4M & 1.4M \\
Hard negatives & Y & N & Y & Y & Y & N & Y & Y & Y \\
Unique targets & 239 & 204 & 239 & 239 & 239 & 181 & 181 & 220 & 220 \\
\bottomrule
\end{tabular}
\end{table}

\section{Chemical Similarity Between Training and Evaluation Sets}
\label{sec:tanimoto-similarity}

A thorough understanding of chemical space coverage and overlap across data used for training and evaluation is paramount in PLI modeling evaluation. In this section we provide analyses of chemical similarity between the \modelname{} training set and evaluation sets. \modelname{} checkpoints are trained on different training set sampling methods. For this analysis we use the training set for checkpoint 0 in \cref{tab:checkpoint-details}, which has the most comprehensive set of target-hit combinations.

\begin{figure}[H]
    \centering
    \includegraphics[width=\linewidth]{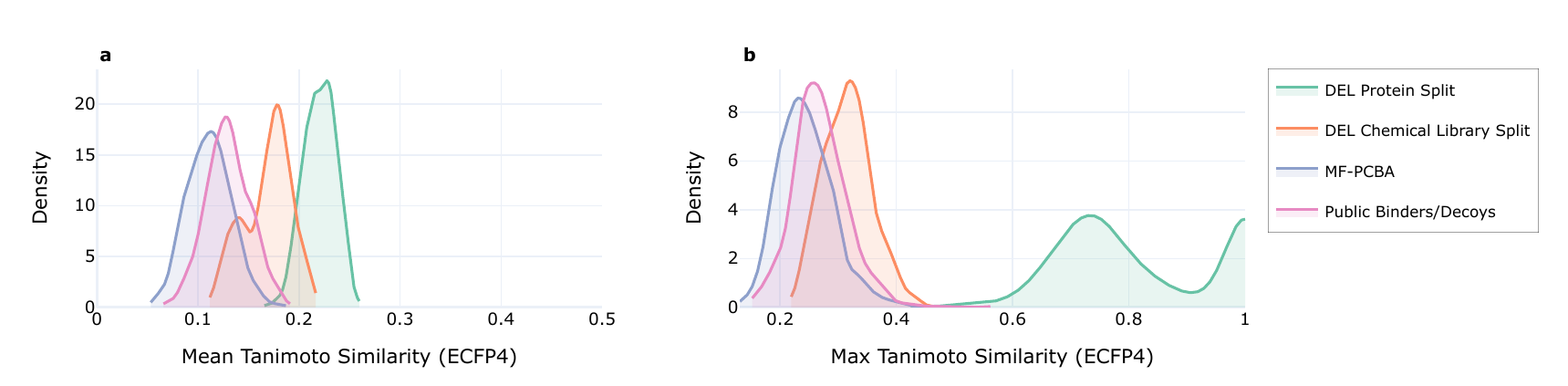}
    \caption{\textbf{Chemical similarity between training and validation sets.} Kernel density estimates of ECFP4 Tanimoto similarity between binders in each validation set and a representative training set. (a) Mean pairwise Tanimoto similarity: for each of 1,000 randomly sampled validation binders, the mean similarity to 10,000 randomly sampled training binders. (b) Maximum nearest-neighbor Tanimoto similarity: for each validation binder, the highest similarity to any hit molecule in the training set. Fingerprints were computed as 2048-bit Morgan fingerprints with radius 2. Note that the $x$-axis scales differ between panels to accommodate the distinct ranges of each metric.}
    \label{fig:tanimoto-similarity}
\end{figure}

\cref{fig:tanimoto-similarity} illustrates the distributions of binder similarities to the DEL training set binders for each evaluation set, showing little structural similarity for three of four evaluation sets. The DEL Protein Split has high similarity with the training set because it was built with DEL screens from the same molecule library, all binders will share at least one, but usually two or all three building blocks with their nearest binder in the training set.

\begin{figure}[H]
    \centering
    \includegraphics[width=\linewidth]{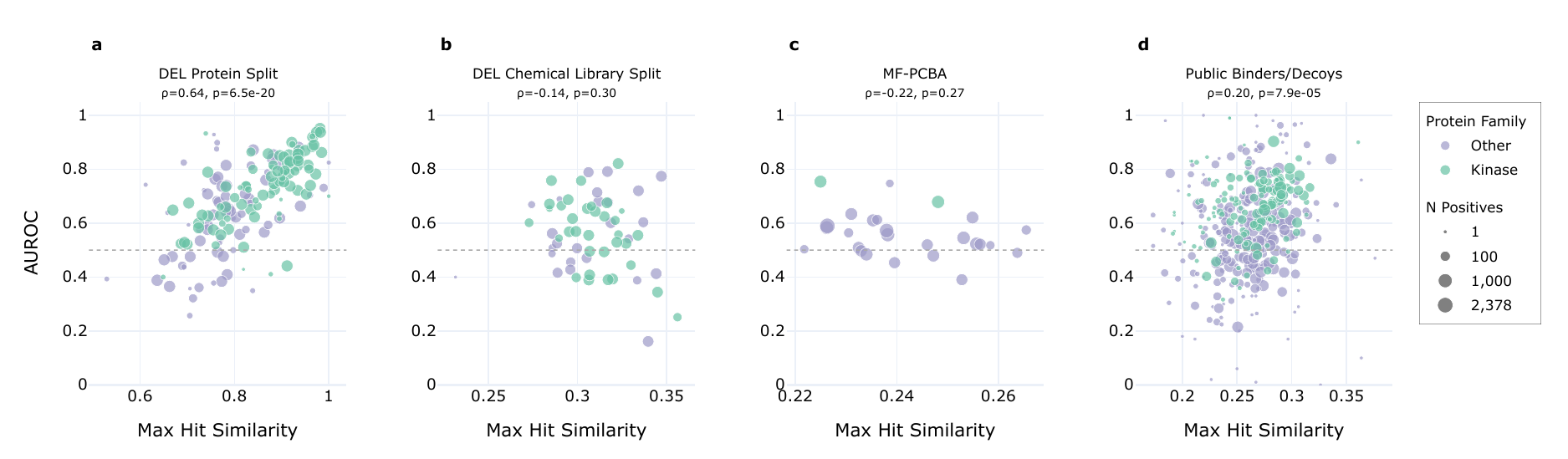}
    \caption{\textbf{Relationship between chemical similarity to training hits and \modelname{} performance across validation sets.} Each point represents a single protein target. The $x$-axis shows the mean maximum (nearest-neighbor) ECFP4 Tanimoto similarity between up to 50 sampled hits for that protein in the validation set and all hits from the training set. The $y$-axis shows per-protein AUROC for \modelname{} predictions. Point size is proportional to the number of positive examples (log-scaled) for each protein, as indicated in the legend. Points are colored by protein family (Kinase or Other). Spearman rank correlation coefficients and $p$-values are shown above each panel. Fingerprints were computed as 2048-bit Morgan fingerprints with radius 2. The dashed horizontal line indicates random-chance AUROC. Note that $x$-axis ranges differ across panels to reflect the distinct similarity distributions of each validation set.}
    \label{fig:similarity-vs-auroc}
\end{figure}

\cref{fig:similarity-vs-auroc} illustrates the relationship between the similarity of the binder structure to training set binders and \modelname{} performance, broken down by protein. The plot reveals a strong relationship for the DEL Protein Split, which is corroborated by the XGBoost Baseline performance (see \cref{fig:xgb-by-hor}), and indicates that binder memorization is an effective strategy for this train/test split. A modest significant relationship is observed for the Public Binders/Decoys set, and no significant effect is observed for either of the remaining datasets.

\section{Benchmark Comparison Details}
\label{sec:benchmark-details}

Benchmark comparison results for \modelname{} vs. Boltz-2 and XGBoost are presented in \cref{sec:benchmark-results}. This section provides additional details.

\subsection{Inference Speed Methodology}

Inference times across all 9 \modelname{} model checkpoints vary significantly based on protein/ligand sequence length and hardware type. We benchmark per-GPU \modelname{} inference time on an 800k-sample subset of the DEL Chemical Library Split validation set at $28.2$ samples/second on H200. For Boltz-2, we use self-reported H100 inference speeds from~\cite{passaro2025boltz2}. While H200 offers approximately $1.2$--$1.5\times$ higher throughput than H100 for transformer workloads, the resulting $\sim 500\times$ hardware-adjusted speedup remains substantial. \modelname{}'s sequence-only architecture enables protein-embedding caching, taking advantage of the low relative protein target diversity in virtually all PLI evaluations.

\subsection{XGBoost Baseline Details}
\label{app:xgboost-baseline}

We trained gradient boosted tree models using XGBoost~\cite{chen2016xgboost} on concatenated protein and ligand representations.

\paragraph{Feature Representation.}
Each protein-ligand pair is represented as a concatenation of protein and ligand feature vectors:
\begin{itemize}
  \item \textbf{Protein features:} We use the CLS token embedding from ESM-2 (650M parameters)~\cite{lin2023evolutionary}, yielding a 1280-dimensional vector per protein sequence.
  \item \textbf{Ligand features:} We compute Extended Connectivity Fingerprints (ECFP4)~\cite{rogers2010extended} using RDKit, with radius 2 and 1024 bits.
\end{itemize}
The final feature vector is 2304-dimensional ($1280 + 1024$).

\paragraph{Training.}
  We train XGBoost classifiers using the histogram-based tree method for computational efficiency. Hyperparameters were tuned through a combination of grid search and manual exploration. The final model used maximum depth 20, minimum child weight 4, gamma 0.4, row subsample ratio 0.5, 50 estimators, and otherwise default parameters. Models were trained on approximately 980,000 protein-ligand pairs.

\begin{figure}[H]
    \centering
    \includegraphics[width=0.4\linewidth]{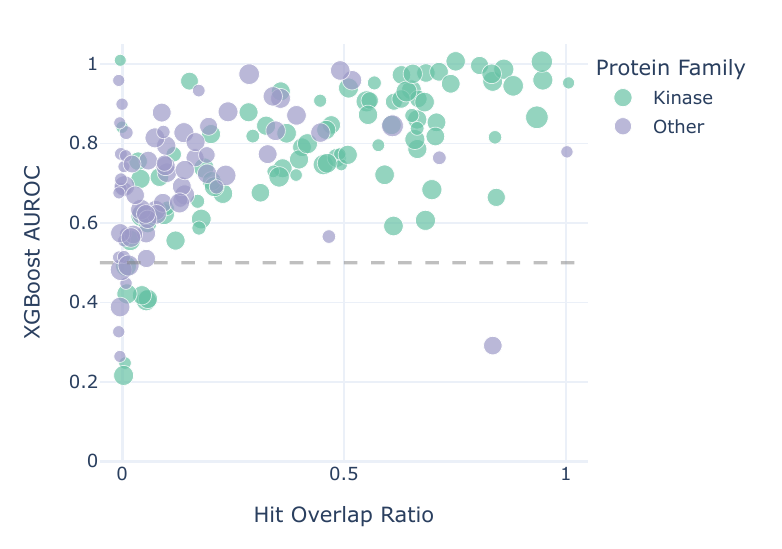}
    \caption{\textbf{XGBoost Baseline per-protein performance by ``hit overlap ratio'',} indicating the proportion of positive samples for a protein target which appear as positive samples against any protein target in the training set. Spearman $r=0.598$.}
    \label{fig:xgb-by-hor}
\end{figure}

\paragraph{XGBoost Baseline vs. DEL Protein Split.}
The XGBoost Baseline achieves substantial performance on the DEL Protein Split, which is composed of protein targets not seen in the training data screened against the same molecule library. This performance reflects the dense nature of large DEL-derived training sets, and the reproducibility and promiscuity of DEL hits against structurally similar targets, especially kinases and kinase sub-families. See \cref{fig:xgb-by-hor}.

\section{Public Dataset Processing}
\label{sec:dataset-processing}

\subsection{Papyrus++}
\label{sec:papyrus-processing}

We construct validation and test sets from the Papyrus++ database~\cite{bequilleux2022papyrus}, a large-scale curated collection of protein-ligand bioactivity data. The raw dataset consists of Papyrus++ v05.5 combined set without stereochemistry, merged with protein target metadata. We apply the following filters: restriction to human proteins, exclusion of proteins with sequence length $> 1024$ amino acids, molecular weight $400 \leq \text{MW} \leq 900$ Da, calculated LogP $0 \leq \text{cLogP} \leq 8$, and rotatable bonds $\leq 12$.

Binding activity labels are derived from mean pChEMBL values using a threshold of 7 (corresponding to approximately 100 nM), a commonly used cutoff for defining active compounds in drug discovery.

To prevent over-representation of heavily assayed protein targets while maintaining chemical diversity, we apply fingerprint-based subsampling for proteins with more than 500 binders. We compute 2048-bit Morgan fingerprints (ECFP4, radius 2) for all binders, calculate the centroid fingerprint, then select $n/2$ molecules closest to and $n/2$ molecules farthest from the centroid in Manhattan distance. This yields a maximum of 250 binders per protein target.

For rigorous evaluation, we construct property-matched decoy sets using molecules from GuacaMol~\cite{brown2019guacamol} as presumed non-binders. For each binder, we identify candidate decoys using $k$-nearest neighbors ($k = 20$) in a standardized descriptor space (MW, Bertz complexity, cLogP, TPSA, number of rings). Candidates are accepted only if their maximum Tanimoto similarity to any known binder for that protein target is $\leq 0.30$, ensuring decoys are chemically distinct from known actives. We generate decoys at a 100:1 ratio, providing a challenging classification task reflective of virtual screening campaigns.

We perform a protein-level split to ensure no protein target appears in both validation and test sets. Validation set performance is used among other dataset performance metrics for \modelname{} model checkpointing.

\subsection{MF-PCBA}
\label{sec:mfpcba-processing}

MF-PCBA is derived from ~\cite{tran2023mfpcba}, a multi-fidelity benchmark sourced from PubChem BioAssay. We filtered the original 60 assays to 28 protein-ligand binding assays, excluding phenotypic screens.

Each assay contains two screening modalities: single-dose (SD) primary high-throughput screening ($\sim$300K compounds per assay) and dose-response (DR) confirmatory screening on SD hits ($\sim$100--1000 compounds). We define binders ($y=1$) as compounds classified as ``Active'' in DR screening, and non-binders ($y=0$) as compounds classified as ``Inactive'' in SD screening; inconclusive DR results are excluded. This labeling strategy ensures binders are confirmed by dose-response validation rather than relying solely on primary screening.

We applied PAINS (Pan-Assay Interference Compounds) filtering using RDKit's \texttt{FilterCatalog} to remove promiscuous compounds exhibiting non-specific activity due to aggregation, redox cycling, or assay interference. Compounds with invalid SMILES representations were also excluded.

The final dataset comprises 28 binding assays across diverse protein targets including proteases, epigenetic modifiers, viral proteins, and metabolic enzymes. Each compound record includes the neutralized SMILES, PubChem CID, SD/DR activity values, and the target protein's amino acid sequence and UniProt accession. Validation and test splits are predefined at the assay level with no protein overlap, enabling evaluation of generalization to novel targets.

\section{Additional Notes}
\label{sec:additional-notes}

\subsection{DEL Chemical Library Split (\strelka{})}
\label{sec:strelka}

\begin{figure}
    \centering
    \includegraphics[width=0.3\linewidth]{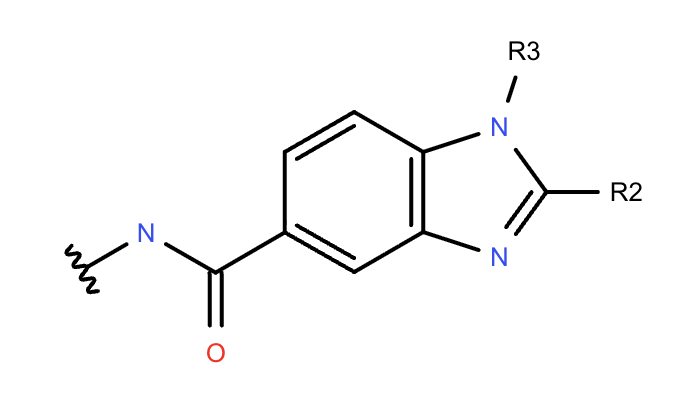}
    \includegraphics[width=0.3\linewidth]{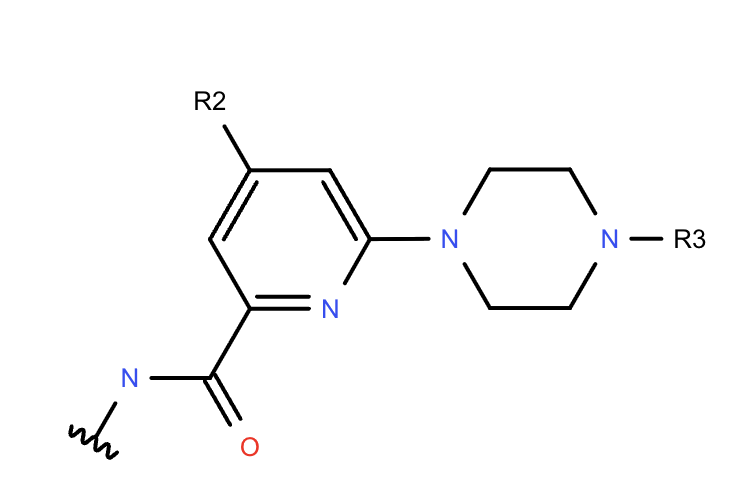}
    \caption{Example Markush structures for (left) molecules in \strelka{} and (right) molecules in the training set ``\delliba{}''}
    \label{fig:placeholder}
\end{figure}

The DEL Chemical Library Split (\strelka{}) benchmark comprises DEL screening data from \dellibb{}, a 1 million member benzimidazole library synthesized by Alphama (Shenzhen, China). \dellibb{} employs a ring-completion chemistry wherein the benzimidazole core is formed during the final synthetic cycle via condensation of \textit{ortho}-nitroanilines with aldehydes—identical to the approach used by \cite{wellaway2020bet} in their discovery of I-BET469, a BET bromodomain inhibitor. Multiple X-ray co-crystal structures from that campaign were deposited in the PDB (6TPX, 6TPY, 6TPZ).

\dellibb{} contains some of the same building blocks used in the GSK library, resulting in exact structural matches between \strelka{} hits and the published benzimidazole series. This provides independent validation of our screening methodology: a library constructed years later by a different organization rediscovered the same BRD4 binders identified in a pioneering DEL-to-candidate campaign.

However, this structural identity is effectively data leakage for structure-based models. Boltz-2 achieves near-perfect performance on BRD4 within \strelka{} (AUROC = 0.986; $n_{\text{pos}}$ = 243, $n_{\text{neg}}$ = 1,262), exceeding its performance on other targets. Since Boltz-2's training data includes PDB complexes, the deposited BRD4-benzimidazole structures constitute direct training examples for molecules in our evaluation set. This case illustrates both the challenge of constructing held-out benchmarks for PDB-trained models and what near-ceiling performance looks like when test compounds overlap with structural training data.

\ifanonym\else
\subsection{Reconciliation with Previously Reported Results}
\label{appendix:blog-reconciliation}

Preliminary \modelname{} performance metrics were presented in an earlier blog post.\footnote{https://leashbio.substack.com/p/good-binding-data-is-all-you-need} The results reported here differ from those initial findings due to several methodological changes, which we document for transparency.

\paragraph{Ensemble expansion.}
The original evaluation employed an ensemble of 3 model checkpoints; the current work uses 9. We observed that validation performance generally improves with ensemble size, and empirically found that 9 checkpoints provided a favorable trade-off between predictive accuracy and inference time.

\paragraph{Training set coverage.}
The blog post reported results on a building-block split of the \delliba{} training set. While building-block splits represent a rigorous evaluation paradigm for DEL-derived data, withholding even a small fraction of building blocks excludes a disproportionately large fraction of enumerated training compounds. The expanded ensemble includes checkpoints trained on the full, unsplit \delliba{} training set, rendering internal building-block splits invalid for the presented models. We therefore leave the evaluation of generalization to novel chemical space to the DEL Chemical Library Split, Public Binders/Decoys, and MF-PCBA benchmarks.

\paragraph{Boltz-2 variability across internal splits.}
We observe a notable discrepancy between Boltz-2 performance on the \delliba{} building-block split and the protein-level split presented in this work. The source of this difference is not fully characterized. PLI binding prediction performance exhibits substantial variation across target space, chemical space, and assay conditions. The building-block split validation set is considerably less diverse in both chemical and target space, which may account for the observed fluctuations, though we cannot offer a definitive explanation.

\paragraph{Negative sampling strategy.}
The blog post additionally reported results on a ``Papyrus Public Validation Set,'' in which binders and non-binders were assigned based on high and low pChEMBL value thresholds rather than synthetic decoys. Subsequent analysis revealed that this labeling strategy introduces substantial bias in basic molecular properties---a known artifact of threshold-based label assignment. For this reason, we shifted to decoy-based negatives for public binding benchmarks.
\fi


\end{document}